# Toward a new mobile cloud forensic framework

Muhammad Faheem
School of Computer Science
University College Dublin
Belfield, Dublin 4, Ireland
muhammad.faheem@ucdconnect.ie

M-Tahar Kechadi, Nhien-An Le-Khac
School of Computer Science
University College Dublin
Belfield, Dublin 4, Ireland
{tahar.kechadi,an.lekhac}@ucd.ie

*Abstract*—Smartphones have created a significant impact on the day-to-day activities of every individual. Now-a-days a wide range of Smartphone applications are available and it necessitates high computing resources in order to build these applications. Cloud computing offers enormous resources and extends services to resource-constrained mobile devices. Mobile Cloud Computing (MCC) is emerging as a key technology to utilize virtually unlimited resources over the Internet using Smartphones. Offloading data and computations to improve productivity, enhance performance, save energy, and improve user experience. Social network applications largely utilize MCC to reap the benefits. The social network has witnessed unprecedented growth in the recent years, and millions of registered users access it using Smartphones. The mobile cloud social network applications introduce not only convenience but also various issues related to criminal and illegal activities. Despite being primarily used to communicate and socialize with contacts, the multifarious and anonymous nature of social networking websites increases susceptibility to cybercrimes. Taking into account, the advantage of mobile cloud computing and popularity of social network applications, it is essential to establish a forensic framework based on mobile cloud platform that solves the problems of today's forensic requirements. In this paper we present a mobile cloud forensic framework that allows the forensic investigator to collect the automated synchronized copies of data on both mobile and cloud servers to prove the evidence of cloud usage. We also show our preliminary results of this study.

*Index Terms*—Mobile device forensics, Mobile cloud forensics, cloud computing, forensic tools.

## I. INTRODUCTION

Smartphones have the capacity to store a broad range of information about the user including e-mail history, location information, usernames, passwords, wireless access point associations and other valuable information [1]. Smartphones took the place of traditional desktops and laptops to access the cloud services and storages using a standard Internet connection. Mobile Cloud technology offers sophisticated applications, which can be used via smartphones from anywhere and at any time [2]. Mobile cloud [3] ensures high scalable and low cost on-demand services to business and information technology organizations. Communications, documents, and multimedia files generated using a mobile device can be moved to cloud-based services for long-term storage. Devices that have the capability of accessing the Internet can offer further connection to online social networks such as Twitter, Facebook, and Skype. Several business organizations have made massive investments in the applications of mobile cloud computing, especially social networking applications [4].

Criminals make use of the cloud services to store and share illicit data related to child exploitation, terrorism, and drug. As the data in the cloud are distributed, there is a great concern for security and trustworthiness [5] and results in multi-jurisdiction issues. The Internet upload and download activities in social networking applications potentially contain forensic-rich data. This information can provide forensic investigators about an individual who is linked to events. Therefore, forensic investigators can find a large amount of data on the cloud that are not available on a mobile device [6]. However, cloud services make difficulties in proving the ownership of a criminal with illicit data as it stores these data in cloud file hosting services.

Cloud storage facilitates the users to store the data and access storage through a variety of internet connected devices [7]. There are several free and paid consumer cloud storage offerings such as Dropbox, Microsoft, SkyDrive, and Google Drive. Remote cloud storage services store the data of the mobile applications like Twitter, Google plus, Facebook, and other data sharing database applications. Any cloud-based mobile applications must ensure data security of the users. It is obvious that the evidence gathered from the mobile and cloud environment has high significance. In recent days, the impact of digital forensics on mobile cloud computing was well exposed. Therefore, it is vital to analyse the cloud-based mobile applications in a forensic manner and collect the data from the corresponding service provider immediately to identify the suspect. In this paper, we show the need of mobile cloud forensics as well as the challenges when we perform forensic acquisition and analysis of mobile cloud application. We then propose a new framework to facilitate the investigation of forensic artefacts from cloud applications on mobile devices. This framework facilitates the forensic capabilities that allow the forensic investigator to collect the automated synchronized copies of data on both the mobile and cloud server to prove the evidence of cloud usage. Therefore, this research establishes a mobile cloud forensic framework that can be extended to any social network application to identify and collect the evidential artifacts in the mobile cloud environment. This work considers only the synced artifacts of mobile and cloud side to prove the mobile's use of cloud service or storage as potential evidential artifacts.

The rest of this paper is organized as follows: Section 2 describes the needs for forensic analysis of mobile cloud social network applications. We discuss on the forensic challenges in Section 3. Next, we present our framework for mobile cloud forensic in Section 4 and we show our preliminary evaluation in Section 5. Finally we conclude and show the future work in Section 6.

## II. NEED FOR FORENSICS IN MOBILE CLOUD SOCIAL NETWORK APPLICATIONS

Social media have witnessed unprecedented growth in the recent years, and billions of registered users access social networks. The forensic investigation in distributed data centres poses several challenges in evidence acquisition and analysis [7]. Mobile users and organization of various business domains such as retail, healthcare, IT and finance stores promote their activities through social network applications and stores data across the clouds using corresponding services. People primarily use mobile cloud social networking applications to communicate and socialize, sharing messages, comments, photographs, audios, and videos, as well as interact with colleagues. The application of mobile cloud computing also includes image processing, natural language processing, GPS and Internet access sharing, applications related to sensor data, querying, and crowd computing [8]. The mobile cloud social network applications introduce not only convenience but also various issues to mobile users. The multifarious and anonymous nature of social networking websites increases susceptibility to cybercrimes. Phishers, child predators, and several other cyber criminals can use the social networking applications with fake identities, concealing their criminal intentions behind innocent-seeming profiles. These innocent-seeming profiles are vulnerable to the danger of secrecy of Android devices [9]. As the data of the mobile devices are stored on a remote server, there is a need forensically to investigate the mobile cloud social network applications.

## III. FORENSIC CHALLENGES IN MOBILE CLOUD SOCIAL NETWORK APPLICATIONS

The main challenges of mobile cloud social network applications lie in collecting the artifacts from mobile device and cloud to establish the evidence by correlating the data. The following are the critical challenges in mobile cloud forensics.

- Applications accessed on mobile devices and executed and stored in the cloud. Most of the existing forensic tools support either cloud or mobile device. Thus, it is necessary for forensic tools and techniques to support mobile cloud-based social network applications.
- Developing forensically sound tool development is a challenging task due to the distributed nature of mobile cloud applications.
- It is difficult to cope up with an updated version of the applications for a forensic analyst to understand features included in the updated version and developing a new strategy for performing forensic analysis with the existing tools.
- Multi-jurisdiction is the major issue because the location of mobile device accessed, and data stored in the cloud are different. In multi-tenancy, the cloud process shared by multiple users in parallel.
- Heterogeneity induces significant complexity in developing forensic tools and methodologies for mobile cloud applications.
- Proving the ownership of the data possessed or accessed by a criminal is a complicated procedure due to the non-uniform authentication and authorization process on device and cloud.
- Generating the required forensic images during the recovery of the remote mobile cloud (application) data is a tedious process. The existing mobile device forensic tools are not suitable for the mobile cloud forensics due to the lack of required capabilities.
- Though, the mobile data stored at a centralized location improves the speed of the forensic analysis, it considers even the unrelated data of the organization for the forensic analysis [10].
- The recovery of forensic rich data on the remote cloud is a complex process if the user clears the sessions.
- One of the major challenges faced with the data stored in cloud services is to identify the exact location of forensic-rich data. It requires more research for identifying and isolating the forensic evidences in the cyber domain, especially in cloud computing. In such cases, mobile devices introduce more difficulty to this cyber domain [11].
- Investigation of mobile cloud application involves several thousands of virtual machines, multiple servers, and several cloud users. Thus, the difficulty lies in detecting a suspect among these users without affecting the privacy of other users [12].
- The forensic investigation in mobile cloud environment compulsorily requires examination of all files in various formats on the disk, handling OS, and file system of each device.

## IV. BASIC OVERVIEW OF PROPOSED METHODOLOGY OF MCFF

This work introduces a novel Mobile Cloud Forensic Framework (MCFF) that supports forensic investigation in social network applications and covers two major phases of digital forensics such as data identification and collection. Any of the social networking applications can extend the proposed MCFF to support the forensic investigation. There are several social networking applications such as Facebook, LinkedIn, Twitter, WhatsApp, and Viber. Some of the common tasks of social networking applications include comment/message postings, uploading (downloading) of photos and videos, and video chatting. These activities leave some kind of traces in both the mobile and cloud environments. This work attempts to prove the evidence of cloud usage using the data collected from the mobile device and through the identification of synced data. The proposed framework takes the advantage of synchronized copies of data on both the device and cloud server to prove the usage of cloud services.

*A. Mobile Cloud Forensic Framework*

This framework can prove the mobile device's use of cloud service through correlating the evidence gathered from mobile and cloud environment. The timeline analysis is the common method of the evidence correlation between various sources. The proposed framework designs a logging framework and deploys in mobile and cloud environment purposely for forensic investigation. The proposed forensic framework facilitates the forensic investigators to conduct forensics even in the absence of either mobile evidential artifacts or cloud evidential artifacts. If a user misuses the social application and uninstalls the application, it is possible to identify this event. Even the uninstallation of the application is recorded in the log in the cloud.

Basically, our proposed framework consists of the following components (Figure 1): mobile device forensics, mobile cloud application forensics, Cloud forensics, Network/wifi forensics and OSINT(Open source intelligence). The correlation of the evidence from all the modules in MCFF and evidence reporting is the final stage of the investigation. In this paper we explore MCFF and mobile device forensics and data acquisition tool that is used for data acquisition by installing it on the target device to acquire basic information i.e. if the phone had encryption, screen saver, passcode, flight mode, developer options enabled or disable, installed apps, contacts, messages, call records etc. The results will be stored on external storage device.

*B. Mobile Device Forensics*

Digital Forensics is the process of inspecting and proving computer crimes in the cyber world [13]. Mobile forensics is a part of digital forensics. Mobile forensics is the process of gathering digital evidence from a mobile device under forensically sound conditions using well-developed tools and techniques. Mobile forensics deals with the process of gathering evidence from a mobile device in forensically sound conditions using well-developed tools and techniques.

The mobile device forensic procedure is divided into stages such as preservation, acquisition, examination and analysis, and reporting [14]. Preservation is the process of seizing and securing suspected mobile devices without modifying the contents of data stored on the devices. Isolation is a technique of evidence preservation and the three common procedures for isolating the mobile device from radio communications are: activate the airplane mode, turn off the device, and finally, place the device in a shielded container [15]. It is proved that placing a device in a shielded container did not provide a complete isolation due to three facts: shielding materials do not ensure enough attenuation, leaks in the shield and conductive shield work as an antenna [16]. Therefore, evidence custodian is advised to use radio isolation techniques to ensure better isolation.

Acquisition is the process of retrieving information from a mobile device and peripheral equipment. Examination and analysis apply forensic tools to discover potential evidence of the mobile device including hidden or obscured evidences. During the examination, if there are any changes in the hash values of two file system extractions, it is necessary to identify the reason for the changes to the files [17]. The data extraction process some way or the other modifies the data [18]. The investigation of the case comes to an end with reporting that maintains a record of all conclusions drawn from the previous phases [14]. Most of the mobile forensic tools are capable of generating automatic reporting. Physical acquisition easily ingresses images of mobile devices into another tool for reporting and permitting analysis of unused file system space while logical acquisition offers a natural and understandable reporting form of acquired information [19].

In our framework, the mobile device forensics is a core component with advanced forensic functionalities. The edge of this component, so-called MCFT, is that it allows investigators perform popular forensic tasks without rooting the devices. Besides, this component also allow to integrate information extracted from OSINT and Cloud platforms to assist the investigation in examining the cloud-based apps on mobile devices.

*C. Cloud Forensics*

Cloud computing is a mutually sharable collection of configurable networked resources that can be reconfigured dynamically with minimal effort [2]. In the cloud environment, the cloud service providers maintain data centers worldwide to guarantee service availability and cost-effectiveness. It replicates the data stored in one data center at several locations to ensure profusion and lessen the risk of failure. The cloud computing offers three levels of services (delivery models) for its clients such as Software as a Service (SaaS), Platform as a Service (PaaS), and Infrastructure as a Service (IaaS) [20]. The cloud has different deployment models and has a direct impact on the geographical location and storage pattern of the data. The deployment models in a cloud include private cloud, community cloud, public cloud, and hybrid cloud [21].

The cloud forensics procedure supports preservation, acquisition, examination, analysis and reporting. During preservation, consider the volatile nature of data while collecting them from cloud sources [22]. The isolation technique preserves the cloud instance and prevents contamination of evidence. Moving suspected instance from one node to another may result in loss of evidence. Therefore, [23] suggested moving other instances in the same node. During acquisition, there is the possibility of losing metadata if the data is downloaded from the cloud [24]. FROST is an acquisition forensic tool designed for the Open Stack cloud platform that supports IaaS cloud and offers the trustworthy forensic acquisition of virtual disks, API logs, and guest firewall logs [25].

Data examination and analysis consumes more time and resources due to the examination of a large number of devices. Several organizations encrypt the data before uploading it to a cloud service [7]. Forensic analyst should be aware of the basic procedures of evidence examination and analysis of the latest cloud computing systems [26]. Hash analysis is a frequently used method to reduce data and refine the investigation-related files [27]. The absence of operating system metadata such as log files, unused space, and temporary files adds more difficulty in a conventional analytical process [28]. The analysis of the client's system may not be sufficient and hence, it recommends analyzing the information stored in the cloud environment by the corresponding user [29]. The presentation of evidences involves in collecting the sources from various contexts.

Cloud forensics is one of the most challenge fields in digital forensics today. As mentioned above, there are different levels of cloud investigation: data centre, infrastructure, platform, virtual machines, wiretap, etc. Hence, our framework does not aim to integrate the cloud forensics components (from third-party, for example) but facilitate users to import forensic artefacts extracted from cloud platforms into the core of our framework where we also apply OSINT techniques (described in the next section) to finally provide enrich information to assist investigators in analysing cloud-based application on mobile devices.

*D. OSINT*

Stands for open source intelligence are tools and techniques which can be used to gather data about person or company. The technique is intended at gathering the possible relation with established with other people in social groups. OSINT can gather further information about suspect using the results obtained from mobile forensics, cloud forensics. Most common tools for OSINT a Focare, LinkedIn Maps, MyiPNeighbours, Google trends, Username search, LeakedIn, Searchcode, Maltego etc. For example we use email and drop box, Review of the emails and phone numbers of other people and entities on mobile. The intelligence on the email is to find the people that the suspect interacts with and contents. Also IP Address- to get the location, the device used, and other contents shared through the IP address.

V. PRELIMINARY EXPERIMENTAL RESULTS

In this section, we present the preliminary results of our framework. These results are from the core component: mobile device forensics. We describe and analyse forensic functions of this component. This component is developed by Android Studio for development so it is native of Android operating system.

*A. MCFT component*

The Data on mobile phones is stored on the SIM (Subscriber Identity Module) card, the internal memory, and the external memory cards. Frequently, there is a large amount of sensitive data stored on mobile phones. However, due to everyday use, it is mixed with huge quantities of irrelevant data. This makes it difficult to retrieve the desired information, and makes the mobile forensics a distinctive and particularly complex branch of forensics.

We have developed forensics tool that is run directly on the Android operating system. Its main function is to perform local forensics through scanning databases without root access. It allows retrieving data from the SIM card and the internal memory, and saving on external memory card.

*B. Local Forensics Functions*

Currently, there is no forensic tool which could directly run on the mobile device. Those that are available need the support of PC Software such as XRY, UFED, OXYGEN or Paraben, etc. The MCFT.apk file is installed directly onto mobile phone from any external storage device or by using adb push command. This file has local evidence browsing functions. This tool scans and retrieves evidence from internal memory such as phone status, device info, applications, messages, call records, SIM details, contacts, configured emails, Wi-Fi history, default browser history and running applications. The MCFT can be used as the first response tool by the law enforcement.

*C. Device Info Function*

This function provides a detailed list of information about the device such as device model, device name, Android version, Android brand, manufacture, Kernel name, Wi-Fi Mac, Wi-Fi SSID, Bluetooth Mac, developer option is enabled or disabled, encryption status, device IMEI, flight mode status, screen lock status, screen saver status, battery status in percentage, as well as date and time configured on the phone.

*D. Phone Status Function*

Phone Status Function is made out of four sub-functions
  a) Screen lock     Enables
If selected, the phone will never be locked and home button or power button awake the phone, user don't have swipe screen to input pin or scan finger.
  b) Screen saver
     If selected, screen saver is disabled
  c) Developer option
     If selected, developer option is enabled
  d) Flight mode
     If selected, flight mode will be switched on.

*E. Configured Emails Function*

This lists all the emails and phone numbers that are registered on the mobile device.

*F. Installed Apps Function*

This function is made out of four parts
  a) All apps
Lists all the apps installed on the phone with date and time of installation.
  b) Third party apps
Lists only third party apps installed by the user. Retrieved data includes time and date of installation.
  c) Disable apps
Lists all the apps that have been disabled by the user.
  d) Uninstall apps
Lists all the apps that have been uninstalled by the user. Retrieved data includes time and date of installation.

*G. View Browser History Function*

This will list default browser history. Data retrieved will include date and time and name of the website which users have accessed.

*H. View Running apps Function*

This will list all the running applications which opened by the user.

*I. View Wi-Fi History*

This will list all the Wi-Fi connections enabled on the device.

*J. SIM Card Functions*

This function shows the status of the SIM card, SIM operator number, SIM Country, Serial number and type

*K. Contact Function*

This function lists all the contacts stored on mobile device.

*L. Messages Function*

This function lists all the messages with time and date of delivery, and a phone number they were received from.

*M. Call record Function*

This function lists all the call records with a phone number, date and time they were received, and whether it was an incoming or outgoing

## VI. CONCLUSION AND FUTURE WORK

MCFT can you used as first response tool by law enforcement. This tool gives a comprehensive summary about the mobile device owner, its applications, system status etc. Mobile device forensic is first step in digital investigation using MCFT.

Advance wireless network, cloud computing and computational power of mobile devices create new opportunities for criminals. Mobile devices forensics is seizing evidence directly from mobile devices but with involvement of cloud computing more and more data need to be seized during communication between cloud and mobile terminal. It means the mobile cloud computing forensics is asked to be done on the mobile devices, at least, the network packages should be captured on the mobile device.

The future work includes building a sniffer application that could catch network packages and analyses the data to do the cloud forensics. We will be also building Open source intelligence tool where we will used the results acquired from mobile devices and cloud forensics.

REFERENCES

[1] Ayers, Richard, "Mobile Device Forensics - Tool Testing", National Institute of Standards and Technology, pp. 1- 23, 2009.
[2] R. Lovell, "White paper: Introduction to cloud computing", ThinkGrid, 2011.
[3] M. Satyanarayanan, "Mobile computing: the next decade," in Proceedings of the 1st ACM Workshop on Mobile Cloud Computing & Services: Social Networks and Beyond (MCS), 2010.
[4] W. Zhenyu, Z. Chunhong, J. Yang, and W. Hao, "Towards Cloud and Terminal Collaborative Mobile Social Network Service," in Proceedings of the 2nd IEEE International Conference on Social Computing (SocialCom), pp. 623, 2010.
[5] Dykstra, J., "Acquiring Forensic Evidence from Infrastructure-as-a-Service Cloud Computing: Exploring and Evaluating Tools, Trust, and Techniques" Digital Forensic Research Workshop (DFRWS), Vol. 43, No. 12, pp.12–19, 2012.
[6] Eoghan Casey and Benjamin Turnbull, "Digital Evidence on Mobile Devices" Elsevier, 2011
[7] Tadjer R. "What is cloud computing?". PCMag.com, 2010.
[8] M. Taylor, J. Haggerty, D. Gresty, R. Hegarty "Digital evidence in cloud computing systems", Digital Investigation, computer law and security review, Vol. 26, pp.304- 308, 2010
[9] Glisson,T. "Calm before the Storm: The Challenges of Cloud Computing in Digital Forensics", International Journal of Digital Crime and Forensics, Vol.4, No.2, pp.28–48, 2012.
[10] Andrew Hoog, "Android Forensics: Investigation, Analysis, and Mobile Security for Google Android"
[11] Mellars, B., "Forensic Examination of Mobile devices" Digital Investigation, Vol.1, No.4, pp.66–72, 2004.
[12] Reiber,L. "SIMs and Salsa, MFI Forum, Mobile Forensics", Vol.13, No.2, pp.15-20, 2008
[13] Finn Ruder. "New study shows 'intent' behind mobile Internet use" Retrieved on 18 February 2012 from: http://www.prnewswire.com/news-releases/new-study-shows-intent-behind-mobile-interetuse-84016487.html, 2012.
[14] Zhu, M, "Mobile Cloud Computing: Implications to Smartphone Forensic Procedures and Methodologies", AUT University, 2011.
[15] Feng Gao, and Ying Zhang, "Analysis of WeChat on IPhone" 2nd International Symposium on Computer, Communication, Control, and Automation (3CA), pp. 278- 281, 2013
[16] Noora Al Mutawa, Ibrahim Baggili, Andrew Marrington, "Forensic analysis of social networking applications on mobile devices" Elsevier transaction on Digital Investigation, Vol. 9, pp. S24–S33, 2012
[17] Levinson, A., Stackpole, B., Johnson, D. "Third Party Application Forensics on Apple Mobile Devices" 44th Hawaii International Conference on System Sciences, pp. 1-9, 2011
[18] Mohammed I. Al-Saleh, and Yahya A. Forihat, "Skype Forensics in Android Devices" International Journal of Computer Applications, Vol. 78, No.7, pp. 38- 44, 2013
[19] Fanlin Meng, Shunxiang Wu, Junbin Yang, and Genzhen Yu "Research of an E-mail Forensic and Analysis System Based on Visualization" 2nd Asia-Pacific Conference on Computational Intelligence and Industrial Applications, 2009M
[20] Buyya, Rajkumar, "Cloud computing and emerging IT platforms: Vision, hype, and reality for delivering computing as the 5th utility" Future Generation computer systems, Vol.25, No.6, pp. 599-616, 2009.
[21] Armbrust, Michael, "A view of cloud computing" Communications of the ACM Vol. 53, No.4, pp. 50-58, 2010.
[22] Ruan, K, Carthy, J & Kechadi, T, "Survey on Cloud Forensics and Critical Criteria for Cloud Forensic Capability: A Preliminary Analysis", ADFSL Conference on Digital Forensics, Security and Law, 2011
[23] M. K. Waldo Delport, Martin S. Olivier, "Isolating a cloud instance for a digital forensic investigation," Springer, Advances in Digital Forensics VIII, International Federation for Information Processing (IFIP) Advances in Information and Communication Technology, Vol. 383, pp 187-200, 2012

[24] Denis Reilly, Chris Wren, Tom Berry, "Cloud Computing: Pros and Cons for Computer Forensic Investigations", International Journal Multimedia and Image Processing (IJMIP), Infonomics Society, Vol.1, No. 1, pp. 26- 34, 2011

[25] Josiah Dykstra, Alan T. Sherman, "Design and implementation of FROST: Digital forensic tools for the OpenStack cloud computing platform" Elsevier transaction on Digital Investigation, Vol. 10, S87–S95, 2013

[26] Taylor, M, Haggerty, J, Gresty, D & Lamb, D, "Forensic Investigation of Cloud Computing Systems", Network Security, No. 3, pp. 4-10, 2011

[27] Hegarty, R, Merabti, M, Shi, Q & Askwith, B, "Forensic Analysis of Distributed Service Oriented Computing Platforms", 12th Annual Post-Graduate Symposium on the Convergence of Telecommunications, Networking and Broadcasting, 2011

[28] G. Grispos, T. Storer, and W. Glisson, "Calm before the storm: The challenges of cloud computing in digital forensics," International Journal of Digital Crime and Forensics (IJDCF), 2012.

[29] Bursztein, E, Cassidian, IF & Martin, M, "Doing Forensics in the Cloud Age Owade: Beyond Files Recovery Forensic", Black Hat, 2011

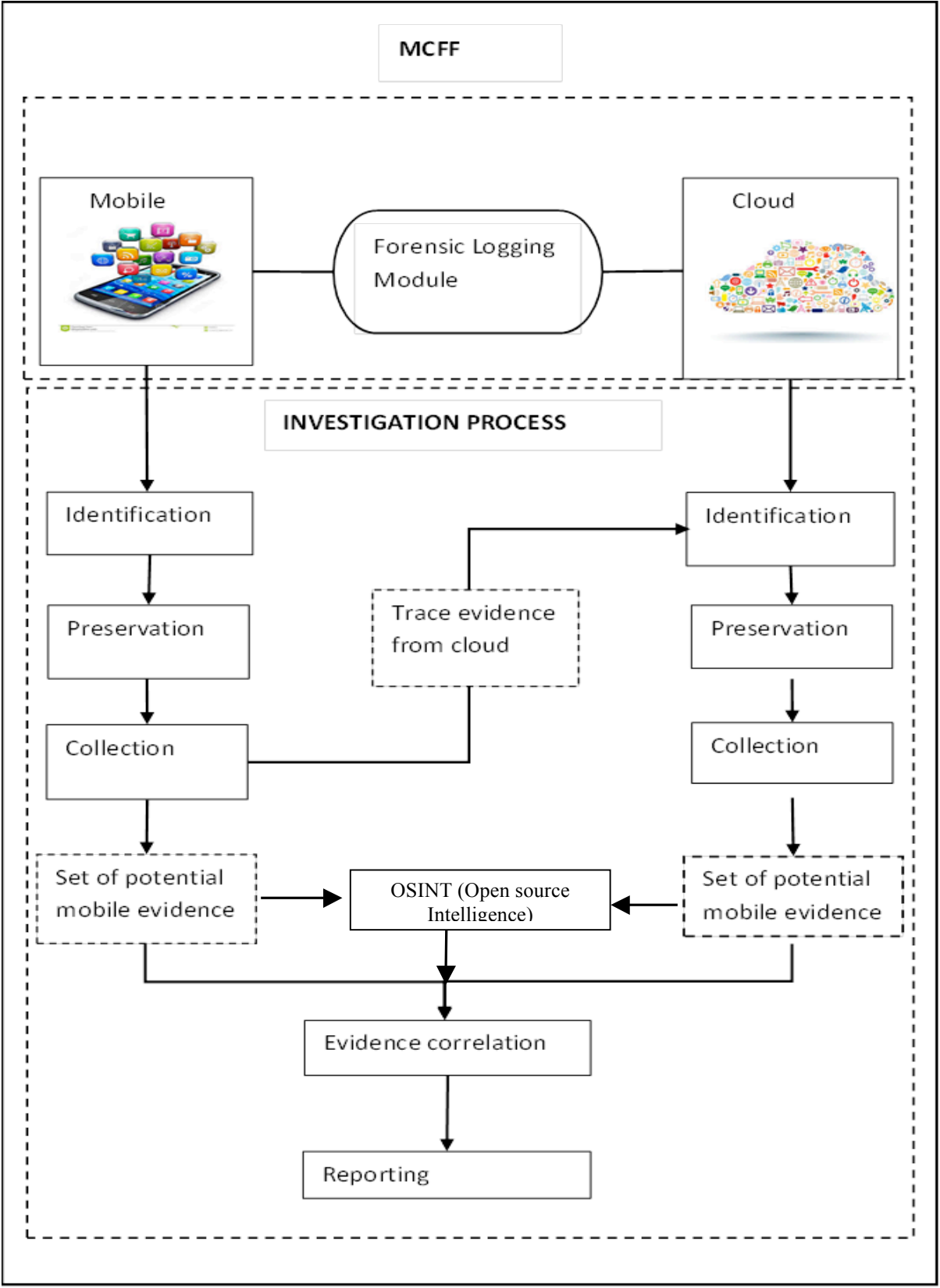

Screen shots of the functions listed in section V

Main page

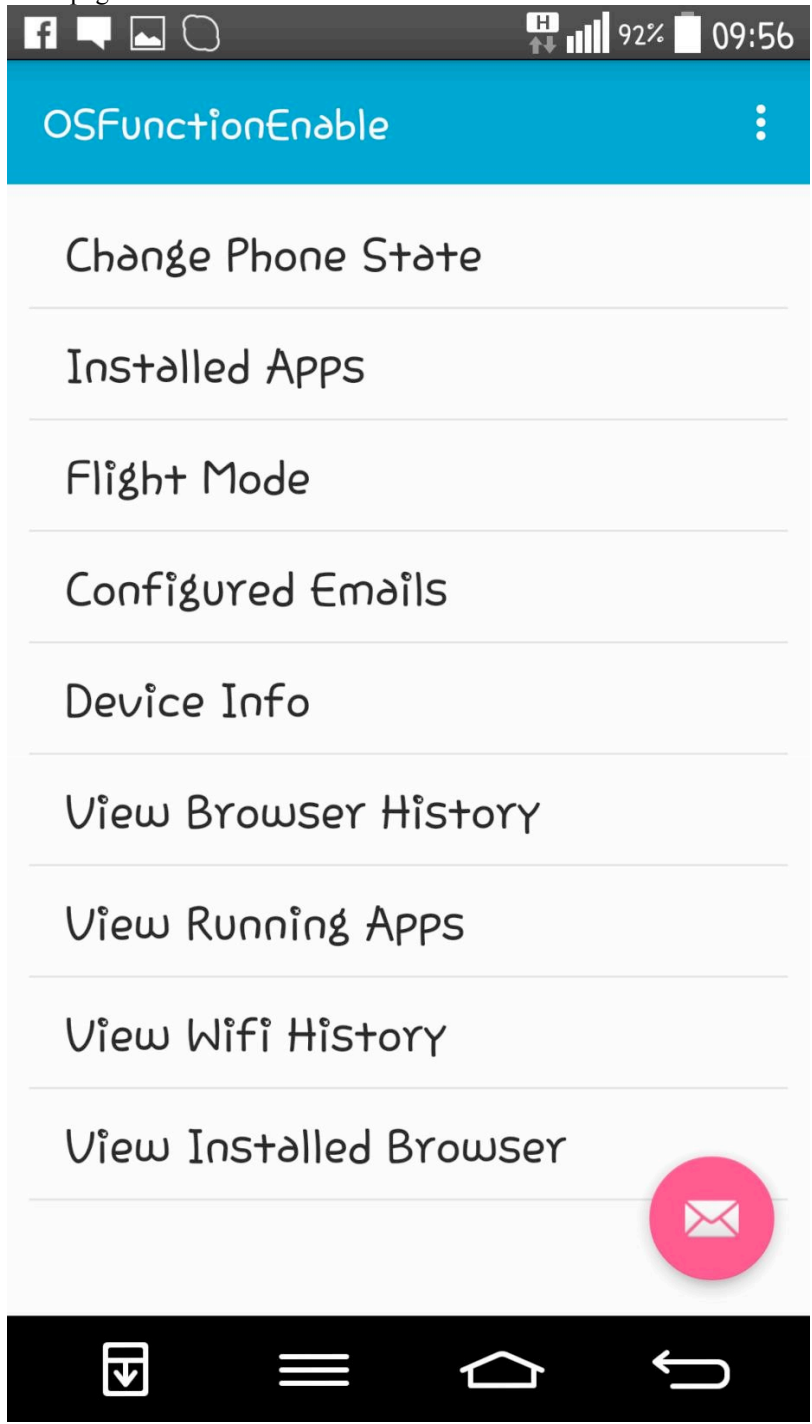

A. Device Info Function

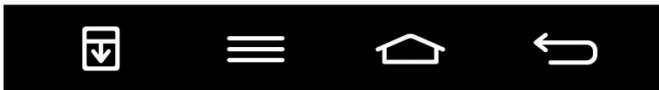

B. Phone Status Function

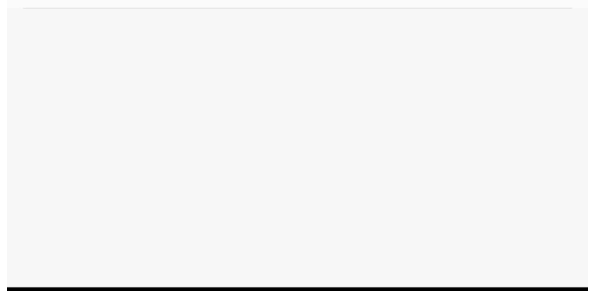
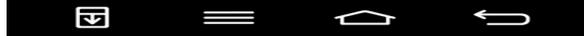

*C. Configured Emails Function*

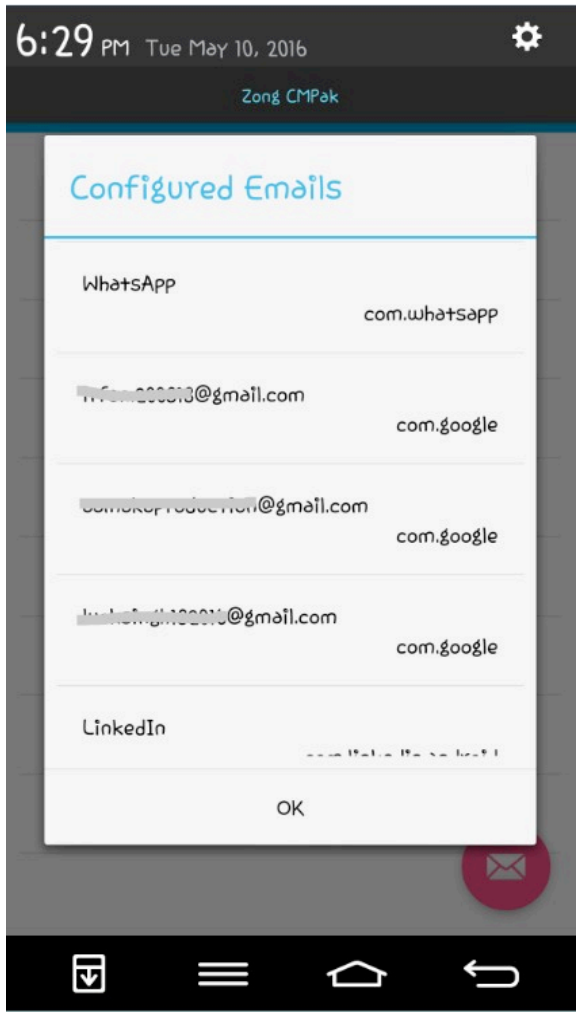

*D. Installed Apps Function*

*E. View Browser History Function*

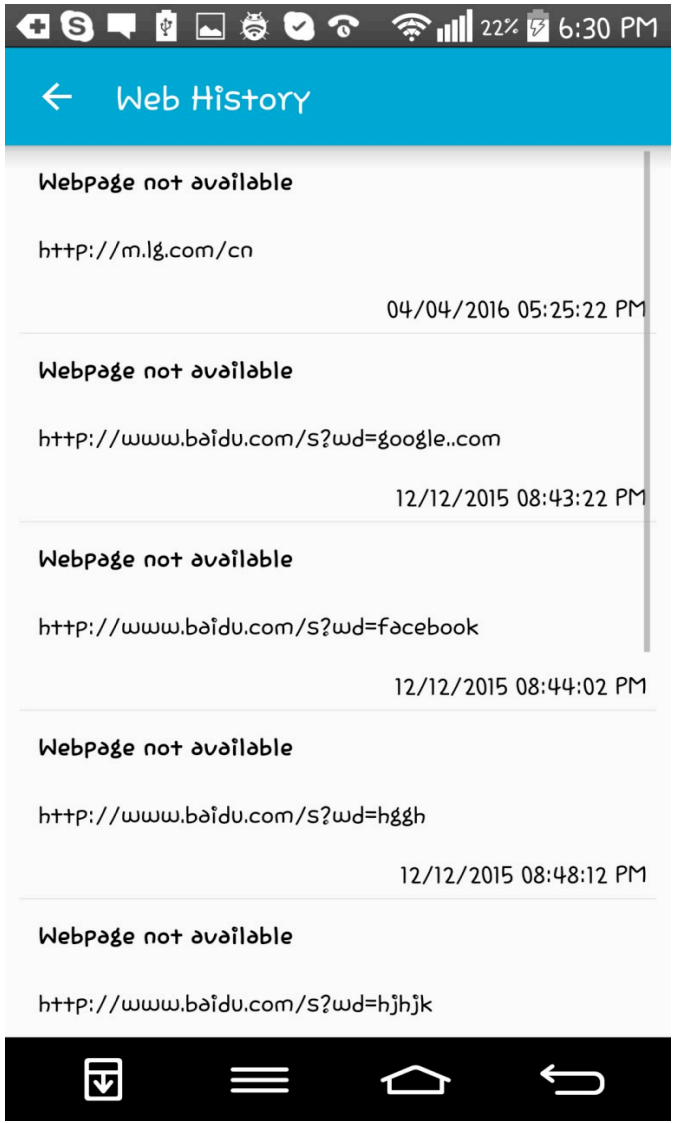

*F. View Running apps Function*

← Running Apps

Home

System UI

Viber

WiFi ADB

TestTest

Google Play Store

Alarm/Clock

Camera

*G. View Wi-Fi History*

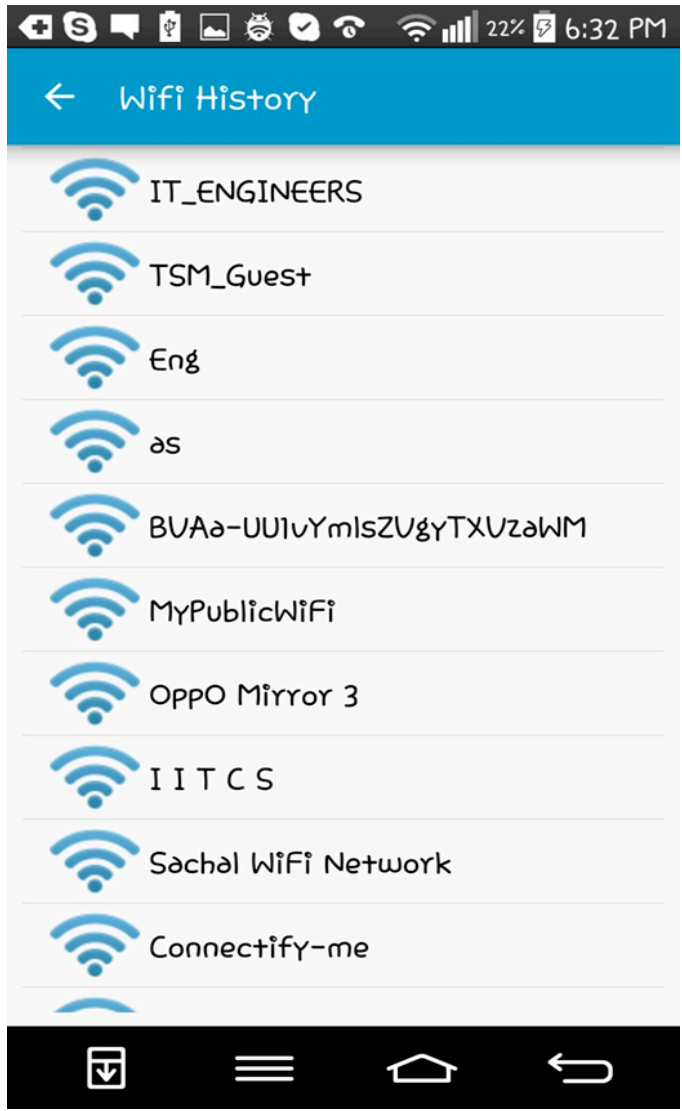